\documentstyle{article}
\begin{document}
\begin{center}
\begin{Large}
\bf{Detection and correction of errors with quantum tomography}\\
\end{Large}
\bigskip
Z. Sazonova\\
Physics Department, Moscow Automobile and Road Construction Institute (Technical University),
64, Leningradskii prospect, Moscow, Russia
\bigskip

R. Singh\\
Wave Research Center at General Physics Institute of Russian Academy of Sciences,
38, Vavilov street, Moscow 117942, Russia \\ Tel./Fax: (+7 095) 135-8234 email: ranjit@dataforce.net \\

\begin{abstract}
It is shown that quantum tomography can detect and correct unlimited number of errors during the evaluation of quantum algorithms on quantum computer.
\end{abstract}
\end{center}
\section{Introduction}
Two main problems of constructing the quantum computer:
\begin{enumerate} 
\item {Quantum decoherence, 
which is inevitable part of quantum computer. Due to the decoherence [16], the 
errors start to appear in quantum algorithms, which destroy the information. 
To remove the errors during the evaluation of quantum algorithms a number 
of error correcting algorithms [7,8,16,19,18,20,21,29,32,34,35] have been discovered. But none of 
them can correct unlimited number of errors. For example, the classical/quantum 
Hamming error-correcting code can correct only one error [7,19,18,27]. Moreover, 
classical/quantum algorithms need additional bits/qubits for the correction 
purposes [7,27]. For example, if we have codeword of three symbols $(n=k=3)$, 
$u=(u_{1},u_{2},u_{3})$. Where $u_{i}$ takes value $0$ or $1$. After 
decoherence the codeword changes the value of symbol from $0$ to $1$ or 
vice-versa. To know whether the codeword $u$ has changed its configuration or not after 
decoherence, we extend the length of codeword from $n=k=3$ to $n=k+1$ or 
$n=k+2$. The $4th$ and $5th$ positions in the codeword, we keep for the 
parity symbols [1,9]. Now the codeword or vector $u=(u_{1},u_{2},u_{3},u_{4},\ldots)$
should satisfy the condition $H.u^{T}=0$ [7,27]. Where $H$ is parity matrix, which 
we choose ourself, so that the condition $H.u^{T}$ should satisfy [7,27].}

\item{ The problems of higher moments. As we know in NMR (nuclear magnetic 
resonance), the higher order moments (total spin $s>1$) are uncontrollable 
or they are not observable with the experimental technique, which we have 
at the moment. It means that the quantum computer, which consists of more 
than $2$ spins of $1/2$ (i.e., total spin $s>1$) always has uncontrollable
moments. So, the density matrix for more than $2$ spins could not reconstructed
fully, with the NMR technique or quantum tomography method experimentally.
But we can reconstruct the density matrix theoretically with the help of 
quantum tomography.}
\end{enumerate}
We will discuss points $1$ and $2$ in more rigorous way in sections $2$, $3$,
$4$, and $5$.

\section{Error-correction with quantum tomography}
The main purpose of error in quantum computer is to flip the phase of 
the quantum state. The state vector of one spin $1/2$ can be written as 
\begin{eqnarray}
|\phi>=a|0>+b|1>.
\end{eqnarray}
Where $|0>$ or $|1>$ stand for the plus or minus direction of the spin 
projection on axis $z$. $a,b$ are the amplitudes of state $|0>$ 
and $|1>$.\\
The decoherence or three Pauli matrices $\sigma_{x}$, $\sigma_{y}$, 
and $\sigma_{z}$ can change the phase and flip the positions of the 
vectors $|0>$ to $|1>$ and $|1>$ to $|0>$ i.e., $\sigma_{x}|\phi>$-
flips the complex amlitudes of the state vector (1), $\sigma_{z}|\phi>$-
changes the phase of the state (1), $\sigma_{y}|\phi>=i\sigma_{x}\sigma_{z}|\phi>$-
changes the pahse and filipping the state vector (1).\\

The same procedure will be for the cases of higher spins, entanglement, and for
mixed states.\\
To correct the quantum errors, we usually use ancilla (additonal) qubits. 
The ancilla qubits contain the error information, which we can extract by 
projecting state vector to the corresponding state [29,32,34,35]. To correct the 
error in one qubit, we need 2 ancilla qubits. It means the total qubits will
be three or total spin $s=s_{1}+s_{2}+s_{3}$ and the uncontrollable moments
will appear in the density matrix, which are not reconstructable by the 
experiment. So, the ancilla qubits will complex our problems more than we 
will have without them, which we will show later.
\section{Problems of higher moments during the reconstruction of density matrix.}
\subsection{Polarization/multipole tensors}
Any square matrix can be decomposed into the polarization tensors [39,6]
\begin{eqnarray}
\hat{A}=\sum_{L=0}^{2\hat{S}}\sum_{M=-L}^{L}A_{L,M}\hat{T}_{L,M}(\hat{S}).
\end{eqnarray}
Where $A_{L,M}=Sp\{{\hat{T}_{L,M}^{+}\hat{A}}\}$ is decomposition coefficient of
matrix $\hat{A}$ and $\hat{T}_{L,M}^{+}(\hat{S})=(-1)^{M}\hat{T}_{L,-M}$ is complex conjugated 
polarization tensor of $\hat{T}_{L,M}$ of total spin $\hat{S}$
\begin{eqnarray}
\hat{T}_{L,M}(\hat{S})=\sqrt{\frac{2L+1}{2\hat{S}+1}}\sum_{m,m^{\prime}}C_{\hat{S},m,L,M}^{\hat{S},m^{\prime}}\chi_{\hat{S},m^{\prime}}\chi_{\hat{S},m}^{+}.
\end{eqnarray}
Where $\chi_{\hat{S},m}$ - spin vector of spin $\hat{S}$ with projection $m$ on $z$-axis and $C_{\hat{S},m,L,M}^{\hat{S},m^{\prime}}$ - is Clebsch-Gordan 
coefficient [6,10,39]. $L$ and $M$ take the values between $0\le{L}\le{2\hat{S}}$ and $=-L\le{M}\le{L}$.

\subsection{Rotation of polarization tensors}
Since the polarization tensors are invariant [39,6] under any rotations. So, we will rotate
our reference frame with Wigner function $\hat{D}(\alpha,\beta,\gamma)$ [10,6,39] with Euler angles 
$\alpha$, $\beta$ and $\gamma$ 
\begin{eqnarray}
\hat{T}_{L,M^{\prime}}^{\prime}=\hat{D}(\alpha,\beta,\gamma)\hat{T}_{L,M}\hat{D}^{-1}(\alpha,\beta,\gamma)=\sum_{M=-\hat{S}}^{\hat{S}}{D^{L}_{M,M^{\prime}}}\hat{T}_{L,M}
\end{eqnarray}
\section{Density matrices}
By using (2), we can decompose density matrix $\hat{\rho}(\hat{S})$ of 
dimensions $n\times{n}$ with decomposition elements $\rho_{L,M}(\hat{S})$ of spin 
$\hat{S}$ i.e.,
\begin{eqnarray}
\hat{\rho}(\hat{S})=\sum_{L=0}^{2\hat{S}}\sum_{M=-L}^{L}\rho_{L,M}(\hat{S})\hat{T}_{L,M}(\hat{S}).
\end{eqnarray}

\subsection{Kronecker product of denisty matrices}
Let the total spin $\hat{S}_{1\ldots{n}}$ is consist of $n$ number of spins i.e., $\hat{S}_{1\ldots{n}}=\hat{S}_{1}+\hat{S}_{2}+\cdots{+\hat{S}_{n}}$. 
The density matrix of a system of $n$ spins is written
\begin{eqnarray}
\label{math/5}
\hat{\rho}_{1\ldots{n}}(\hat{S}_{1\ldots{n}})=\otimes_{i=1}^{n}\hat{\rho}_{i}(\hat{S}_{i}).
\end{eqnarray}
Where $\otimes_{i=1}^{n}\hat{\rho}_{i}(S_{i})=\underbrace{\underbrace{\underbrace{\hat{\rho}_{1}(\hat{S}_{1})\otimes{\hat{\rho}_{2}}(\hat{S}_{2})}_{1^{st}}\otimes{\hat{\rho}_{3}}(\hat{S}_{3})}_{2^{nd}}\otimes{\cdots}\otimes{\hat{\rho}}_{n}(\hat{S}_{n})}_{n^{th}}$
The $1^{st}$ underbrace denotes the kronecker product of $\hat{\rho}_{1}(\hat{S}_{1})$ and $\hat{\rho}_{2}(\hat{S}_{2})$, the $2^{nd}$ denotes the kronecker product of the result of $1^{st}$ underbrace and $\hat{\rho}_{3}(\hat{S}_{3})$ and so on.
By using (5), we can decompose (6) into
\begin{eqnarray}
\hat{\rho}_{1\ldots{n}}(\hat{S}_{1\ldots{n}}) & = & \sum_{L_{1},L_{2},\cdots,L_{n}=0,0,\cdots,0}^{2\hat{S}_{1},2\hat{S}_{2},\cdots,2\hat{S}_{n}}\sum_{M_{1},M_{2},\cdots,M_{n}=-L_{1},L_{2},\cdots,L_{n}}^{L_{1},L_{2},\cdots,L_{n}}\{[(\hat{\rho}_{L_{1},M_{1}}\nonumber \\
 & & \otimes{\rho}_{L_{2},M_{2}}(\hat{S}_{2}))\otimes{\rho}_{L_{3},M_{3}}(\hat{S}_{3})]\otimes{\cdots}{\otimes}{\rho}_{L_{n},M_{n}}(\hat{S}_{n})\}\{[(\hat{T}_{L_{1},M_{1}}(\hat{S}_{1}) \nonumber \\
 & & \otimes{\hat{T}_{L_{2},M_{2}}}(\hat{S}_{2}))\otimes{\hat{T}_{L_{3},M_{3}}(\hat{S}_{3})]\otimes{\cdots}\otimes{\hat{T}_{L_{n},M_{n}}(\hat{S_{n}})}}\}.
\end{eqnarray}
The kronecker product [6,10,39] of two polariation tensors $\hat{O}_{L_{1},M_{1}}(\hat{S}_{1})$ 
and $\hat{O}_{L_{2},M_{2}}(\hat{S}_{2})$ having spins $\hat{S}_{1}$ and $\hat{S}_{2}$ can be written as
\begin{eqnarray}
\label{math/7}
\hat{O}_{L_{1},M_{1}}(\hat{S}_{1})\otimes{\hat{O}_{L_{2},M_{2}}(\hat{S}_{2})}=\sum_{\hat{S}_{12}=|L_{1}-L_{2}|}^{L_{1}+L_{2}}\sum_{M=-\hat{S}_{12}}^{\hat{S}_{12}}C_{L_{1},M_{1},L_{2},M_{2}}^{\hat{S}_{12},M}\hat{O}_{L,M}(\hat{S}_{12}).
\end{eqnarray}
By applying the kronecker product property (8), we can get the density matrix (7) of dimensions $n\times{n}$ in the following way: first we
apply (8) to density matrices $\hat{\rho}_{1}(\hat{S}_{1})\otimes{\hat{\rho}_{2}(\hat{S}_{2})}$ and then the kronecker product of result of $\hat{\rho}_{1}(\hat{S}_{1})\otimes{\hat{\rho}_{2}}(\hat{S}_{2})$
and $\hat{\rho}_{3}(\hat{S}_{3})$ and so on.
\subsection{Examples in quantum information theory}
First of all, we will consider the density matrices for two spins $\hat{S}_{1}$, $\hat{S}_{2}$ and then
gradually increase the $n$ number of spins upto $3$. Since, we are interested in quantum information theory, so
we are considering all the spins with spin $1/2$.
\subsubsection{Density matrix of two spins}
By using (3), (5), (6), (7), and (8), we can write density matrix $\hat{\rho}_{12}$ of 
two spins $\hat{S}_{1}$ and $\hat{S}_{2}$
\begin{eqnarray}
\hat{\rho}_{12}(\hat{S}_{12}) & = & \hat{\rho}_{1}(\hat{S}_{1})\otimes{\hat{\rho}_{2}(\hat{S}_{2})}\nonumber \\
 & = & \sum_{L_{1},L_{2}=0,0}^{2\hat{S}_{1},2\hat{S}_{2}}\sum_{M_{1},M_{2}=-L_{1},-L_{2}}^{L_{1},L_{2}}[\rho_{L_{1},M_{1}}(\hat{S}_{1})\otimes{\rho_{L_{2},M_{2}}(\hat{S}_{2})}]\nonumber \\
 & {} & [\hat{T}_{L_{1},M_{1}}(\hat{S}_{1})\otimes{\hat{T}_{L_{2},M_{2}}(\hat{S}_{2})}]\nonumber \\
 & = & \sum_{L_{1},L_{2}=0,0}^{2\hat{S}_{1},2\hat{S}_{2}}\sum_{M_{1},M_{2}=-L_{1},-L_{2}}^{L_{1},L_{2}}\sum_{\hat{S}_{12}=|L_{1}-L_{2}|}^{L_{1}+L_{2}}\sum_{M_{12}=-\hat{S}_{12}}^{\hat{S}_{12}}\nonumber \\
 & {} & C_{L_{1},M_{1},L_{2},M_{2}}^{\hat{S}_{12},M_{12}}C_{L_{1},M_{1},L_{2},M_{2}}^{\hat{S}_{12},M_{12}}
[\rho_{\hat{S}_{12},M_{12}}(\hat{S}_{12})][\hat{T}_{L_{12},M_{12}}(\hat{S}_{12})].
\end{eqnarray}
$\hat{T}_{S_{12},M_{12}}$ can be defined from (3). The spin vectors $\chi$ for $\hat{S}_{12}$ can be defined by diagonalizing the spin Hamiltonian.

\section{Problems in the reconstruction of density matrix}
The problems arise when we want to reconstruct some elements of density matrix for spin 
$\hat{S}>1$. That is, we could not reconstruct all the elements of density matrix 
experimentally. Theoretically, there are no problems of reconstruction of density matrix
elements. The higher order moments [23] are responsible for the reconstruction of density
matrix for $\hat{S}\ge{1}$. We will discuss later about the moments, which can be detected with experiment and 
which not.
\subsection{Power expansion of electrostatic and magnetostatic field}
If the distance between the two charged particles (nuclei) is large as compare 
to their dimensions then the electrostatic and magnetostatic field between 
them can be expanded into power series [23,24,30]
\subsubsection{Power expension of electrostatic field}
\begin{eqnarray}
\varphi^{(l)}_{electro}=\frac{1}{R^{(l+1)}_{0}}\sum_{m=-l}^{l}\sqrt{\frac{4\pi}{2l+1}}Q_{m}^{(l)}Y_{l,m}^{\ast}(\Theta,\Phi)\\
Q_{m}^{(l)}=\sum_{a}e_{a}r_{a}^{(l)}\sqrt{\frac{4\pi}{2l+1}}Y_{l,m}(\Theta_{a},\Phi_{a})
\end{eqnarray}
Where $e_{a}, r_{a} -$ electric charge and radius of $a^{th}$ nucleus
relative to its adjacent nucleus. $\vec{R}_{0} -$ radius vector of all nuclei
to the point of observation. $Q_{m}^{(l)} -$ tensor [5,6,10,11,14,31,39]
of rang $l$ with $2l+1$ independent components. $Y_{l,m}^{\ast}(\Theta,\Phi) -$ 
complex conjugated spherical harmonic of angles $(\Theta$,$\Phi)$ betweeen 
vectors $\vec{R}$, $\vec{r}$, and coordinate axes.

By using the spherical harmonic properties [24,14,39], we can express 
spherical harmonics into eigenvectors of integer spin $\hat{S}$. For example, 
for $\hat{S}=1$, we have
\begin{eqnarray}
Y_{1,0}=\sqrt{\frac{3}{4\pi}}\cos(\Theta)=
\left(
\begin{array}{c}
0\\
1\\
0
\end{array}
\right)\\
Y_{1,1}=-{\sqrt{\frac{3}{8\pi}}\sin(\Theta)\exp(i\Phi)}=
\left(
\begin{array}{c}
0\\
0\\
1
\end{array}
\right)\\
Y_{1,-1}={\sqrt{\frac{3}{8\pi}}\sin(\Theta)\exp(-i\Phi)}=
\left(
\begin{array}{c}
1\\
0\\
0
\end{array}
\right)
\end{eqnarray}

\subsubsection{Power expension of magnetostatic field}
The power expension of magnetostatic field can be written similarly as in the
case of electrostatic field, i.e.,
\begin{eqnarray}
\varphi^{(l)}_{magnetic}=\frac{1}{R^{(l+1)}_{0}}\sum_{m=-l}^{l}\sqrt{\frac{4\pi}{2l+1}}Q_{m}^{(l)}Y_{l,m}^{\ast}(\Theta,\Phi)\\
Q_{m}^{(l)}=\sum_{a}e_{a}r_{a}^{(l)}\sqrt{\frac{4\pi}{2l+1}}Y_{l,m}(\Theta_{a},\Phi_{a})
\end{eqnarray}

\subsection{Electric quadrupole moments}
Nuclei with spins $\hat{S}=1/2$ have symmetric distribution of their charges. So,
the elerctric quadrupole moments have value $Q=0$. Nuclei with spins $\hat{S}\ge{1}$ 
have asymmetric distribution of charges which create electric quadrupole moment 
$Q\ne{0}$. The $Q$ moment can interact with gradient of electric field at nucleus. 
To find the electric quadrupole moments of a system of spins, we have two options: 
1. By applying gradient of electric field. 2. Through acoustic nuclear magnetic 
resonance.
\subsection{Longitudinal $(T_{1})$ and transverse $(T_{2})$ relaxations}
Spin-spin interactions, spin-lattice interaction are responsible for relaxation 
times $T_{1}$ and $T_{2}$.

\subsection{Quantum tomography of spin systems}
Theoretically, we can reconstruct the density matrix by applying the quantum 
tomography to spin systems. It means, we have to rotate density matrix with 
Euler angles $(\alpha,\beta,\gamma)$ and take integration over $\alpha,\beta,\gamma$,
which is not possible experimentally.\\
Examples: 

\subsection{Higher order moments}
As the number of spin increases, the higher order moment problems arise.\\
For example:\\
For the case of two spins $1/2$ ($\hat{S}_{1}=1/2$, $\hat{S}_{2}=1/2$ and $\hat{S}=\hat{S}_{1}+\hat{S}_{2}$),
we have magnetic dipole moment and electric quadrupole moment [5,6,10,11,31,39], which can be defined from the experiment.\\
For the case of three spins $1/2$ ($\hat{S}_{1}=1/2$, $\hat{S}_{2}=1/2$, $\hat{S}_{3}=1/2$ and 
$\hat{S}=\hat{S}_{1}+\hat{S}_{2}+\hat{S}_{3}$), we have magnetic dipole moment, 
electric quadrupole moment, magnetic quadrupol moment and electric octoploe moment. 
The higher order moments: magnetic quadrupole moments, electric octople moments are not
extractable from the experiment due to the absence of comlicated experiment scheme.\\
For the case of higher number of spins $1/2$. The higher order moments appear, which 
are undefined from the experiment except magnetic dipole moment and electric quadrupole 
moments.

\subsubsection{Spin Hamiltonian for $2$ spins $1/2$}
Hamiltonian of two spins $\hat{\sigma}_{1}$ and $\hat{\sigma}_{2}$ defined in
linear space $S_{1}, S_{2}$. $\hat{\sigma}_{1z}$, $\hat{\sigma}_{2z}$ coupling 
with hyperfine interaction $J_{12}$ are placed parallel to applied constant 
magnetic field $B_0\|$$z-axis$. For simplicity, we are taking $\hbar=1$.
$$
\hat{H}2=-\mu{B_0}\cdot{(\hat{\sigma}_{z1}\otimes{\hat{E}_2}})-\mu{B_0}\cdot{(\hat{E}_{1}\otimes{\hat{\sigma}_{z2}}})+J_{12}(\hat{\sigma}_{x1}\otimes{\hat{\sigma}_{x2}})+J_{12}(\hat{\sigma}_{y1}\otimes{\hat{\sigma}_{y2}})
$$
\subsubsection{Spin Hamiltonian for $3$ spins $1/2$}
Hamiltonian of three spins $\hat{\sigma}_{1}$, $\hat{\sigma}_2$ and $\hat{\sigma}_{3}$ 
are defined in linear space $S_{1}, S_{2}, S_{3}$ and coupling with hyperfine interaction 
$J_{12}$ between $\hat{\sigma}_{1}$ and  $\hat{\sigma}_{2}$, $J_{23}$ between $\hat{\sigma}_{2}$
and $\hat{\sigma}_{3}$ and $J_{31}$ between spins $\hat{\sigma}_{3}$ and
$\hat{\sigma}_{1}$. $\hat{\sigma}_{1z}, \hat{\sigma}_{2z}$ and $\hat{\sigma}_{3z}$ are placed 
parallel to applied constant magnetic field $B_0\|$$z-axis$:

\begin{eqnarray*}
\hat{H}3=-\mu{B_0}\cdot{(\hat{\sigma}_{z1}\otimes{\hat{E}_2}}\otimes{\hat{E}_3})
-\mu{B_0}\cdot{(\hat{E}_{1}\otimes{\hat{\sigma}_{z2}}}\otimes{\hat{E}_{3}})\nonumber \\
-\mu{B_0}\cdot{(\hat{E}_{1}\otimes{\hat{E}_{2}}}\otimes{\hat{\sigma}_{z3}})
+J_{12}(\hat{\sigma}_{1}\otimes{\hat{\sigma}_{2}}\otimes{\hat{E}_{3}})\nonumber \\
+J_{23}(\hat{E}_{1}\otimes{\hat{\sigma}_{2}}\otimes{\hat{\sigma}_{3}})
+J_{31}(\hat{\sigma}_{3}\otimes{\hat{\sigma}_{1}}\otimes{\hat{E}_{2}})
\end{eqnarray*}
Hamiltonians of higher number of spins $1/2$ can be written in the same way as for $2$
and $3$ spins $1/2$.

\subsection{Kronecker product in quantum information theory to get
the spin Hamiltonians}
To write the spin Hamiltonian, first of 
all we should write $\hat{S}_{x}$, $\hat{S}_{y}$, $\hat{S}_{z}$, $\hat{S}^{2}$ and 
then add them, we will get the spin Hamiltonians (e.g. $\hat{H}2$ and $\hat{H}3$) in the following way\\
Let we want to write the spin Hamiltonian of $n$ nuclear spins in NMR (Nuclear Magnetic Resonance):\\ 
\begin{enumerate}
\item{Total projection of spins on z-axis is conserved.}
\begin{eqnarray}
\hat{S}_{z}=1/2(\hat{\sigma}_{1z}
\left\{
\begin{array}{cc}
\otimes^{n}_{i=2}\hat{E}_{i}, & \mbox{if } n\ge{2}\\
1 & \mbox{if } n=1.
\end{array}
\right\}
+\hat{E}_{1}\otimes{\hat\sigma}_{2z}
\left\{
\begin{array}{cc}
\otimes^{n}_{i=3}\hat{E}_{i}, & \mbox{if } n\ge{3}\\
1 & \mbox{if } n=2\\
0 & \mbox{if } n<2.
\end{array}
\right\}\nonumber \\
+\hat{E}_{1}\otimes{\hat{E}}_{2}\otimes{\hat{\sigma}_{3z}}
\left\{
\begin{array}{cc}
\otimes^{n}_{i=4}\hat{E}_{i}, & \mbox{if } n\ge{3}\\
1 & \mbox{if } n=3\\
0 & \mbox{if } n<3.
\end{array}
\right\}
+\cdots )
\end{eqnarray}
$\hat{S}_{x}$ and $\hat{S}_{y}$ can be written by putting 
$\hat{\sigma}_{x}$ and $\hat{\sigma}_{y}$ in place of $\hat{\sigma}_{z}$.
\item The square of the total spin
$
\hat{S}^2=\hat{S}\cdot{(\hat{S}+1)}
$ is conserved.
\begin{equation}
\hat{S}=1/2(i\hat{\sigma}_{x}+j\hat{\sigma}_{y}+k\hat{\sigma}_{z})\nonumber \\
\end{equation}
\begin{equation}
\label{math/100}
\hat{S}^2=1/4(\hat\sigma_{x}^{2}+\hat{\sigma}_{y}^{2}+\hat{\sigma}_{z}^{2})
\end{equation}
\item Equations (14) and (3) are constant of motion.
That is $[\hat{S}_{z},\hat{S}^{2}]=0$. It means that the eigenvalues and 
eigenvectors of (14) and (3) are identicals.
\item The Hamiltonians $\hat{H}2$ and $\hat{H}3$ are consist of two parts,
addition of (14) and (3). So the eigenvalues and 
eigenvectors of equations (14) and (3) are also the eigenvalues and 
eigenvectors of Hamiltonians $\hat{H}{2}$ and $\hat{H}{3}$.
\end{enumerate}
\section{Conclusion}
We propose an alternative technique (quantum tomography), by using that 
we can correct the errors and can reconstruct the higher moments. So, 
the ancilla qubits make the error-correction problem more complicate. To
remove the errors in quantum computer, we need to develop the experimental 
techniques how to detect higher moments.

\end{document}